\documentstyle[aps,prl,psfig,preprint]{revtex}
\begin{document}
\tightenlines
\draft
\preprint{}
\title{Superinsulator Phase of two-dimensional Superconductors}
\author{A.~Kr\"amer and S.~Doniach}
\address{Dept.~of Applied Physics, Stanford University, Stanford CA 94305,
U.S.A.}
\date{July 22, 1998}
\maketitle
\begin{abstract} 
Using path-integral Quantum Monte Carlo we study the low-temperature phase diagram
of a two-dimensional superconductor within a phenomenological model,
where vortices have a finite mass and move in a dissipative environment
modeled by a Caldeira-Leggett term. The quantum
vortex liquid at high magnetic fields exhibits superfluidity and thus 
corresponds to a {\em superinsulating} phase which 
is characterized
by a nonlinear voltage-current law for an infinite system 
in the absence of pinning. This superinsulating phase is shifted to higher
magnetic fields in the presence of dissipation.
\end{abstract}
Quantum fluctuations of vortices play an important role
for the low-temperature 
physics of two-dimensional (2D) superconductors. It has been established
that  
dissipative \cite{Bla-Ivl}, inertial \cite{Mag-Cep,Ono-Don}, or Hall-type \cite{Roz-Str} 
zero point motion is able to melt the vortex lattice at sufficiently
high magnetic fields and give way to a quantum vortex liquid phase
in principle at $T=0$.
In the presence of disorder this melting transition is a continuous one
and corresponds to 
the localization or crystallization of Cooper pairs in the theory of the 
superconductor-insulator transition \cite{Fis,Don-Inu}.
However, for a system {\em without disorder}, the melting of the 
vortex lattice is discontinuous in (2+1) dimensions\cite{Ono-Don}.

In this letter, using
path-integral Quantum Monte Carlo (QMC), we will 
 study  a phenomenological model where the vortices
have a finite mass and  move in a dissipative environment,
which is modeled by a Caldeira-Leggett term. 
Both, vortex
mass and dissipation are assumed to originate from electronic contributions
while the vortices themselves are considered as bosons. 
For the simulations we focus on the situation where the Magnus force is 
zero which is appropriate for the case of granular systems
or Josephson Junction arrays (JJA) where experiments suggest that the
transverse force is very small \cite{granular}. The effect
of a weak Magnus force will then be discussed afterwards.

In a thin film superconductor at high magnetic fields vortices effectively
interact logarithmically.  So the question arises whether the vortices
can form a superfluid or not. 
We will show using QMC that a 2D system of 
logarithmically interacting particles indeed exhibits superfluidity
at low temperatures and high densities even in the presence of 
dissipation.\cite{remark}
This means that  the quantum vortex liquid actually corresponds to a 
{\em superinsulating} phase  in which an
infinitesimal current will cause  an infinite voltage in
the absence of pinning effects
in analogy with sliding charge density
waves. 
This superinsulating phase exhibits a non-linear voltage-current 
law with non-universal exponent and is different from the insulating 
Cooper pair glass phase in Fisher's \cite{Fis} treatment of the 
field-tuned superconductor-insulator transition in the presence of
strong disorder. We will show numerically that at finite 
temperature the superinsulator gives way to a classical vortex 
liquid via a Kosterlitz-Thouless (KT-) transition. In the presence of
dissipation
this phase transition is shifted to higher magnetic fields 
possibly giving rise to an intermediate non-superfluid vortex liquid phase at 
$T=0$.

In the non-dissipative case the vortex system is described by the
Euclidean Lagrangian
\begin{equation}
\label{model}
{\cal L}=
\frac{m}{2}\sum\limits_k\left(\frac{d{\bf r}_k}{d\tau}\right)^2
-T_0\sum\limits_{k>l}\ln(|{\bf r}_k-{\bf r}_l|),
\end{equation}
where ${\bf r}_k(\tau)$ are the vortex positions, $m$ is the vortex
mass, and $T_0$ controls the strength of the logarithmic interactions. 
We assume the system to be embedded in a neutralizing background 
which corresponds to the presence of a uniform external magnetic field 
$B=\Phi_0\varrho$ so that 
the long-range interaction in Eq.~(\ref{model})
is well-defined even for periodic boundary conditions.
Here, $\varrho$ is the vortex density, and
$\Phi_0$ is the flux quantum.
${\bf r}_k(\tau)$ obeys
periodic boundary conditions in imaginary time $\tau$, that include 
bosonic exchange of the particles.

Finite-temperature QMC simulations 
have been performed for systems with up to 128 time slices and 
N=16, 28 and 36 particles on a grid of 256$\times$222 sites 
in a rectangular periodic
box which was commensurable with a triangular vortex lattice. 
A  bisection algorithm was used where several particles are updated 
in several imaginary time slices simultaneously in order to allow for 
cutting and reconnection of the paths \cite{Binder}.
The superfluid density $\varrho_s$ was then obtained from
the distribution of winding numbers $W_i$, ($i=x,y$) around the periodic 
cell by \cite{Pol-Cep}
\begin{equation}
\label{winding}
\frac{\varrho_s}{\varrho}
=\frac{mL_i^2}{\beta\hbar^2N}\left< W_i^2 \right>,
\end{equation}
where $L_i$ is the periodic box length in $i$-direction, 
and $\beta$ is the inverse temperature. 
For the lowest-temperature data reported in this work typically about $10^5$ sweeps were needed to equilibrate the winding
number moves. Equilibration was checked by carefully
monitoring the distribution of $W_i$ which can be fitted by a Gaussian at
low $T$.
In the following we use dimensionless variables, where the temperature 
$T$ is given in units of $T_0$, 
and the magnetic field $B$ is given in units
of $B_0=(mT_0\Phi_0)/(\sqrt{3}\;\hbar^2)$. 
 
The phase diagram of the model (\ref{model}) is shown in Fig.~1a.
It consists of
three phases: a superconducting vortex solid phase (VS) at low densities and temperatures, a
classical vortex liquid (VL) at high temperatures, and a quantum vortex
liquid 
(QVL) at
high densities and low temperatures. 
The vortex solid thermally melts via a KT-transition mediated by
the unbinding of dislocation pairs which takes place at $T_{KT}\approx
0.0071$ \cite{Hub-Don} and is approximately independent of the magnetic field
$B$. At low temperatures quantum fluctuations melt the vortex solid via a
discontinuous transition at the melting field $B_M\approx 0.0077$ 
\cite{Mag-Cep} which is approximately independent of the temperature
\cite{Ono-Don}.
In order to check our numerical algorithm we measured the internal energy
at various fields and low temperature, and also estimated $B_M$ by
calculating the Bragg-peak intensity. The results where found to be
in good quantitative agreement with the zero temperature results reported 
in Ref.~\cite{Mag-Cep}.

The inset in Fig.~1a shows the superfluid fraction $\varrho_s/\varrho$ as 
function of
the magnetic field $B$ at $T=0.005$ as obtained in our simulations.
At the quantum melting transition $\varrho_s/\varrho$ jumps from zero
to one which quantitatively reproduces the result of Nordborg and Blatter
\cite{Nor-Bla}, and shows that the QVL actually is a vortex superfluid with
$\varrho_s=\varrho$.
In the following we will focus on the transition from the QVL phase
to the VL phase.
The behavior of the superfluid fraction  is shown in  
Fig.~2a where $\varrho_s/\varrho$ is plotted as function of the temperature $T$ 
for different magnetic fields $B$. It is
seen
that all data collapse onto one single curve when plotted against the
scaling
variable $T/B$. This scaling behavior  
can be understood with a KT-type 
unbinding of topological (vortex-antivortex) excitations (which we will call {\em dual} vortices)
in the vortex-superfluid. Since the energy scale of the logarithmic
interaction of dual vortices is given by
$T_0'=2\pi\hbar^2\varrho/m$, the transition temperature thus scales with $B$.
Because of the fundamental duality between vortices and charges \cite{duality}  the 
dual vortices are actually related to the Cooper pair degrees of freedom 
in the system as will be discussed below.

Following Ceperley and Pollock \cite{Cep-Pol} further numerical evidence 
for the existence of a KT-transition can be obtained by 
performing a finite-size analysis that explicitly invokes the
KT-recursion
relations \cite{Kos-Tho} which are integrated up to the system radius $L/2$. There
are only two independent fit parameters for all numerical data involved in 
this procedure: the dual vortex core energy $E_0$ and the dual vortex
radius  $d$.
The resulting fits are the solid curves in Fig.s 2a and 2b. Fig.~2b 
shows data for 
$N=$16, 28, 36, and also extrapolated to $N=\infty$. From this the
KT-transition temperature is obtained as
$T_{KT}'=1.451\;\hbar^2\varrho/m$, where $E_0=6.49\;\hbar^2\varrho/m$, and
$d=0.634\;\varrho^{-1/2}$.
Interestingly it is observed that the core size $d$
scales with the vortex spacing $\varrho^{-1/2}$ which is a consequence of the 
logarithmic interactions which provide no additional length scale.

In order to interpret  these numerical results we note that the vortex 
superfluid actually should be considered as a {\em charged} superfluid where the flux quantum $\Phi_0$
plays the role of a ``charge'' and the ``flux quantum'' of the dual vortices
is given by $2e$. In a homogeneous superconductor the corresponding 
gauge field ${\bf a}$ is related to the 2D-superfluid density of Cooper
pairs $n_s$, $\nabla\times {\bf a}=2e n_s$,
which gives rise to the Magnus force acting on the vortices 
\cite{Ao-Tho}. 
To be consistent with the assumption 
of of a zero transverse force in a granular system we therefore suppose 
that  $n_s$ is replaced by the density of
fluctuating  charges $\delta n_s$ on the superconducting islands,
$\nabla\times {\bf a}=2e \delta n_s$, which is a usual assumption in the
context of granular superconductors or JJA's \cite{Delft}. 
Static fluctuations  of the gauge field $\bf a$ can approximately be
included
by a contribution $(\Lambda/4e^2)(\nabla\times {\bf a})^2$ to the free energy
of the vortex superfluid, where 
$\Lambda$ is the strength of the ($\delta$-) interaction between the
Cooper pairs \cite{remark-EC}. The numerical simulations then correspond to
the case of strong repulsion, $\Lambda=\infty$, where $\delta n_s$ decays
logarithmically around a dual vortex. 
For finite $\Lambda$ 
the vortex-superfluid current flowing around the dual vortices is screened on
the length $\lambda=\left(2\Lambda m/\hbar^2\varrho\right)^{1/2}$.
In the following we assume that this screening length is much larger
than the dual vortex core size $d$.

Let us discuss the dynamics of the vortex superfluid in the case 
$\Lambda=\infty$. Since dual vortex excitations carry a charge $\pm 2e$
the separation of dual vortex-antivortex pairs leads to an 
electric current $J$. However, in an infinite system 
this process requires an infinite amount of energy when the electric
field $E$ is zero, while for a non-zero electric field the energy  barrier
is finite. It is well known \cite{V-I} that for $T\ll T_{KT}'$ 
this type of dynamics
leads to a non-linear voltage-current law with nonuniversal exponent, 
 \begin{equation}
\label{V-I}
E\sim J^{T/T_0'}.
\end{equation}
Eq.~(\ref{V-I}) explicitly shows that the vortex superfluid phase
actually is a
{\em superinsulating} phase with infinite resistivity even at 
non-zero temperature.
In a realistic system finite-size
effects as well as a finite screening length $\lambda$ will restore a finite resistivity
$\rho\sim e^{E_0/T}$ for small currents where the Arrhenius factor 
$e^{-E_0/T}$ controls the number of
dual vortex-antivortex pairs in the system, and 
the KT-transition  will be broadened. 

Let us now briefly consider the case where a weak  Magnus force is present.
This means, that the vortices on average pick up a phase $\phi=2\pi n_s^0
\xi_0^2\neq 0$ when encircling a superconducting grain of area $\xi_0^2$
\cite{Zhu-Ao}. Here, $2en_s^0$ is the charge density of Cooper pairs 
contributing to the transverse force which plays the role of a fictitious 
external magnetic field. By analogy with an ordinary superconductor we
therefore expect that dual vortices will form an Abrikosov lattice.
Thermal excitations  and dynamics of this
dual vortex lattice  will then be governed by the unbinding of  
{\em dislocation} pairs and 
the KT-melting temperature is given by $T_{KT}''\approx
0.06\;\hbar^2\varrho/m$.
This again is a superinsulating phase described by the non-linear 
voltage-current law (\ref{V-I}), 
where $T_0'$ is replaced by the interaction energy 
of dislocations. The corresponding phase diagram is schematically shown
in Fig.~1b.

In what follows we will discuss the influence of 
dissipation, which is treated approximately by adding a
Caldeira-Leggett term (time-delay) \cite{Cal-Leg}
\begin{equation}
\label{CL}
{\cal L}_{CL}=
-\frac{\eta}{2\pi}\sum\limits_k
\int\limits_0^{\hbar\beta}d\tau'
\frac{d{\bf r}_k}{d\tau} 
\ln\left|\sin\frac{\pi}{\hbar\beta}(\tau-\tau')\right| 
\frac{d{\bf r}_k}{d\tau'}
\end{equation}
to the Lagrangian (\ref{model}). Eq.~(\ref{CL}) 
phenomenologically describes the coupling of each single vortex to a
separate harmonic oscillator heat bath which is characterized by 
a frictional constant $\eta$. This model is appropriate to describe
dissipation
originating from normal currents in the vortex cores \cite{Bla-Ivl} and
has also been used by Wheatley \cite{Wheatley} to study the statistics
of holons coupled to the background of spinons in the context of
high-$T_c$ superconductivity. 
With regard to superfluidity there is one  essential assumption 
in using Eq.~(\ref{CL}) to model a dissipative environment. 
The coupling to independent
harmonic oscillator heat baths cannot exactly be valid for
indistinguishable particles. Following \cite{Wheatley} we therefore assume
that dependencies between separate heat baths induced by bosonic exchange
of the vortices can be neglected.

We again computed the superfluid fraction which
is shown in Fig.~3 for $B=0.04$, $0.08$, and $\eta=0.01$,
$0.02$, where $\eta$ is given in units $\eta_0=\pi mT_0/\hbar$.
It is seen that dissipation leads to a shift 
of the KT-transition to lower temperatures and, more surprisingly, to the 
suppression of superfluidity at $T=0$. This reentrance behavior can be
understood by 
the contribution of the Caldeira-Leggett term (\ref{CL}) to the action of vortices
involved
in a particle  exchange, which can be described in terms of a 
temperature-dependent effective mass, $m^\ast=m+\beta\hbar\eta(\ln 2/\pi)$.
For the bare vortex superfluid density $\tilde\varrho_s$ (unrenormalized by dual
vortex-antivortex pairs)  we therefore make the ansatz 
$
\tilde\varrho_s=\varrho (m/m^\ast),
$
which is a linear function of $T$ at 
low temperatures.
Since the interaction of dual vortices scales
with $\tilde\varrho_s$, the KT-transition is then
shifted to lower temperatures, $T_{KT}'^{,\eta}=T_{KT}'-(\eta\hbar/m)(\ln
2/\pi)$. 
Using this ansatz and taking the core-energy to be $E_0^\eta=E_0\tilde\varrho_s/\varrho$ we perform the same
finite-size fit as above which yields the solid curves in Fig.~3 and 
shows a remarkable coincidence with the numerical data. Again the same
values
for $E_0$ and $d$ as before were used so that no new fit parameters appear
here. The dashed curves in Fig.~3 show the superfluid fraction
extrapolated for
$N=\infty$. At low temperature, $\varrho_s/\varrho$ rises linearly with
$T$
 and reaches its maximum value $\approx 1-0.15\,\eta/(\hbar\varrho)$ near $T_{KT}'^{,\eta}$.

The KT-transition line between the VL and the QVL
(extrapolated to $T=0$)  is shown in the $B$-$T$-diagram in 
the inset of Fig.~3
where the friction parameter $\eta$ was assumed to be independent of the
temperature. The crossing point with the $B$-axis ($T_{KT}'^{,\eta}=0$) 
is given by 
$B_{KT}^0 \approx 0.15\, \eta\Phi_0/\hbar$. If $B_{KT}^0\gtrsim B_M$ this 
suggests the existence of an intermediate field range
in which the classical vortex liquid phase continues to $T=0$.\cite{remark-shift}
In this regime the superinsulating phase is unstable against
the unbinding of dual vortex-antivortex pairs even at $T=0$ since
the strength of the logarithmic dual vortex-vortex
interaction is proportional to $T$. This
also means that the exponent of the voltage-current law (\ref{V-I}) will stay 
finite in the limit $T\rightarrow 0$.

We add a few comments about the experimental accessibility of the results
presented in this paper. The notion of a quantum melting transition as it
was treated here requires a system with weak pinning as well as the existence 
of a finite vortex mass which has
to be small enough in order to push the melting field $B_M$ significantly
below the upper critical field $B_{c2}$. 
Indeed, a field-tuned
superconductor-insulator transition which can be related to the melting
of the vortex lattice has been observed in
Josephson junction arrays\cite{Zant-et-al}. According to our results for a finite system
the conductivity in the superinsulating phase is expected to show
an Arrhenius behavior $\sigma\sim e^{-E_0/T}$ where dissipation 
(as it was modeled here) leads to a linear $T$-dependence of 
$E_0$ at low temperatures so that the conductivity 
stays finite in the limit $T\rightarrow 0$. This is qualitatively  consistent
with the experimental results in Ref.~\cite{Zant-et-al}.

In conclusion, using QMC simulations, we have studied 
superfluidity in a system of logarithmically interacting massive vortices 
moving in a dissipative environment. 
It was shown that  the quantum vortex liquid at
high magnetic fields 
corresponds to a {\em superinsulating} phase characterized by a non-linear 
voltage-current law with non-universal exponent. 
This superinsulating phase gives way to a classical vortex-liquid
phase via a Kosterlitz-Thouless transition, which is
shifted  
to higher magnetic fields in the presence of dissipation. 
\acknowledgments
We thank D.~Das for useful discussions. We also gratefully acknowledge 
financial support by the Deutsche Forschungsgemeinschaft and the NSF 
under grant No.~DMR 9627459. Numerical work has partly been done at
the San Diego Supercomputer Center (SDSC). 

\begin{figure}
\centerline{\psfig{figure=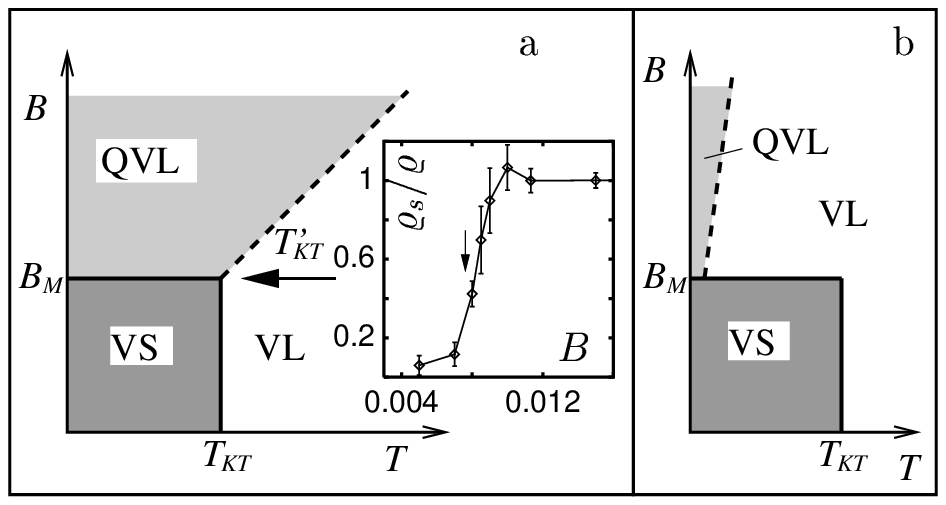,width=8.3cm}}
\caption{($B$-$T$)-phase diagram  of the 
vortex system in a 2D superconductor (below $B_{c2}$). 
The inset shows the jump in the superfluid density at the quantum melting
transition at $B_M\approx 0.0077$ (arrow).
(a) zero Magnus force. (b) weak Magnus force.}
\end{figure}
\begin{figure}
\centerline{\psfig{figure=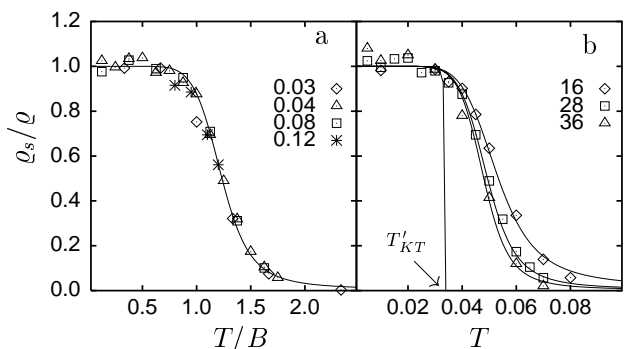,width=8.3cm}}
\caption{Superfluid fraction $\varrho_s/\varrho$: (a) 
as function of the scaling variable $T/B$ for the magnetic fields
$B=$ 0.03, 0.04, 0.08, and 0.12, and a 28-particle system. (b) at  $B=0.04$
as function
of $T$ for systems with 16, 28, and 36 particles and extrapolated
for an infinite system.}
\end{figure}
\begin{figure}
\centerline{\psfig{figure=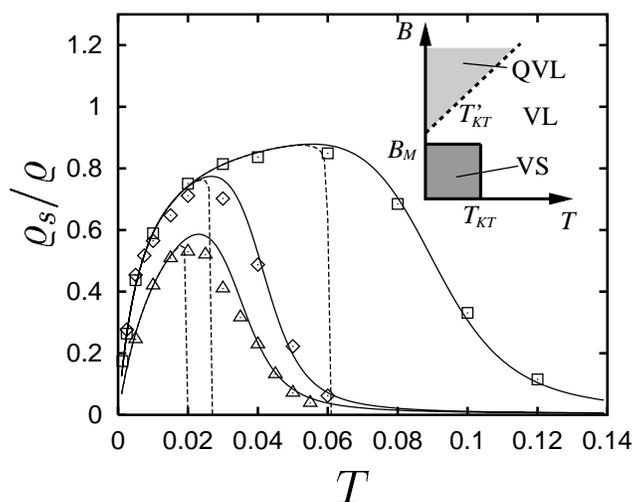,width=8.3cm}}
\caption{Superfluid fraction $\varrho_s/\varrho$ as function
of $T$ for non-zero dissipation ($N=28$): $B=0.04$, $\eta=0.01$ ($\Diamond$),
$B=0.04$, $\eta=0.02$ ($\triangle$), and $B=0.08$, $\eta=0.01$ ($\Box$).
Dashed curves: Finite-size extrapolation to $N=\infty$. Inset:
($B$-$T$)-phase diagram (schematically).}
\end{figure}


\begin{references}
\bibitem{remark} Recently Nordborg and Blatter \cite{Nor-Bla} 
have done a similar 
calculation for studying the entanglement of 3D classical vortices
at high temperatures.
\bibitem{remark-EC} In a JJA the parameter $\Lambda$ is related to the
charging energy $E_C$.
\bibitem{remark-shift} We note that a non-zero friction constant $\eta$ also
shifts the melting field $B_M$ itself to higher values. However, our
simulation results indicate that this 
shift is smaller than the shift of the superinsulating phase.
\end{references}
\end{document}